\documentclass[sigconf,nonacm]{acmart}
\AtBeginDocument{%
  \providecommand\BibTeX{{%
    \normalfont B\kern-0.5em{\scshape i\kern-0.25em b}\kern-0.8em\TeX}}}

\setcopyright{acmlicensed}
\copyrightyear{2024}
\acmYear{2024}
\acmDOI{XXXXXXX.XXXXXXX}

\acmConference[FDG '24]{Foundations of Digital Games 2024}{May 21--24,
  2024}{Worcester, MA}
\acmISBN{978-1-4503-XXXX-X/18/06}

\begin{document}

\title[Cheap and Easy Open-Ended Text Input for Interactive Emergent Narrative]{Cheap and Easy Open-Ended Text Input\\for Interactive Emergent Narrative}

\author{Max Kreminski}
\email{mkreminski@scu.edu}
\affiliation{%
  \institution{Santa Clara University}
  \city{Santa Clara}
  \country{USA}
}


\begin{abstract}
We present a demonstration of Play What I Mean (PWIM): a novel, AI-supported interaction technique for interactive emergent narrative (IEN) games and play experiences. By assisting players in translating high-level gameplay intents (expressed as short, unstructured text strings) into concrete game actions, PWIM aims to support open-ended player input while mitigating the overwhelm that players sometimes feel when confronting the large action spaces that characterize IEN gameplay. In matching player intents to game actions, PWIM makes use of an off-the-shelf sentence embedding model that is lightweight enough to run locally on a player's device, and wraps this model in a simple user interface that allows the player to work around occasional classification errors. 
\end{abstract}



\begin{teaserfigure}
    \centering
    \includegraphics[width=\textwidth]{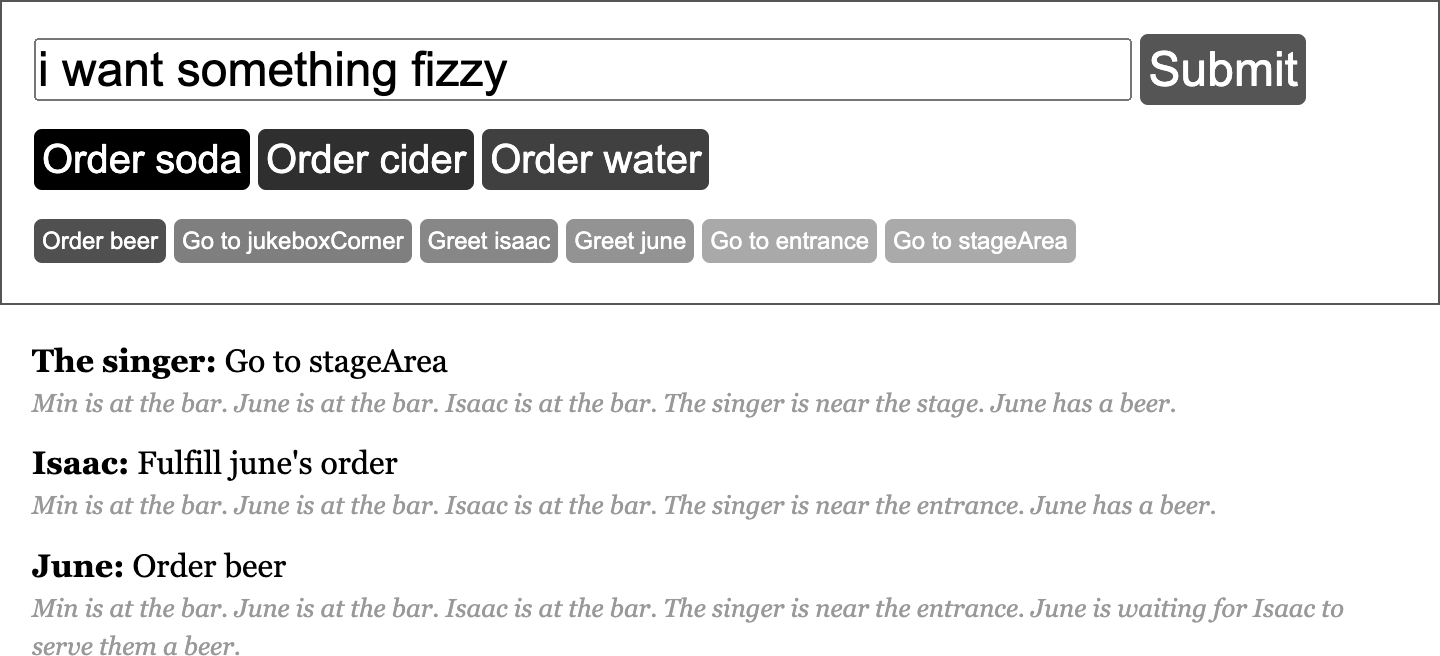}
    \caption{A screenshot of our demo of PWIM. The text area in the top left allows the player to describe what they want to do; possible actions are presented below as clickable buttons, colored and sized to highlight actions that are similar to the input.}
    \label{fig:PWIM}
\end{teaserfigure}


\maketitle

\section{Play What I Mean}
Interactive emergent narrative (IEN) games and play experiences, such as \emph{Dwarf Fortress} and \emph{The Sims}, allow players to perform actions in a wide-open sandbox that then responds narratively to whatever the player has done~\cite{LouchartEmergentNarrative,IENHistory}. 
The wide-open nature of player input in IEN supports many of the form's central aesthetics~\cite[Sec. 3.2]{CuratingSimulatedStoryworlds}, but also tends to lead to player \emph{overwhelm}~\cite{Overwhelm} at the sheer breadth of actions that can be performed. Avoiding overwhelm while retaining open-endedness of input thus stands out as a central design challenge for IEN games.

Overwhelm at the breadth of the action space in IEN is partly downstream of user interface design: a graphical UI that makes all available game actions visible to the player, when backed up by a very wide action space, tends to appear visually overwhelming. For instance, the Versu UI~\cite{VersuJournalArticle}---which displays all available actions as distinct, clickable buttons---is sometimes described as confusing, due to the very large number of actions that must be explicitly considered at each choice point.

Alternatively, interactive narrative games with wide action spaces have often made use of a text parser to interpret commands entered by the player as short text strings, allowing players to access the full breadth of a wide-open action space without navigating an overwhelming GUI. But parsers have problems of their own: verbs in parser interactive fiction are notoriously hard to discover~\cite{GuessTheVerb}, the syntax to which parser-interpreted instructions must conform is notoriously finicky~\cite{InfocomTypeParser}, and the implementation of a truly forgiving parser for open-ended utterances (such as that in \emph{Fa\c{c}ade}~\cite{Facade}, which aims to respond to arbitrary player-entered dialogue) represents a major investment of development time and effort---so major that very few games have ever shipped with such a parser.

To address these difficulties, we demonstrate a new interaction technique for IEN that aims to provide some of the advantages of open-ended text input without necessitating the development or reuse of a sophisticated parser. This technique---Play What I Mean (PWIM)---builds on the basic \emph{text-to-dialogue} approach proposed by \citeauthor{TextToDialog}~\cite{TextToDialog} (i.e., using a sentence embedding model to map player-entered intent strings to game-provided lines of player character dialogue) by combining it with a more sophisticated, Versu-inspired system for state tracking and conditional action provision~\cite{Praxish}, as well as a user interface that enables players to explicitly choose between several game actions that represent \emph{likely} matches for their intent. Our demo of PWIM can be played in a web browser\footnote{\url{https://mkremins.github.io/pwim}} and is implemented completely in clientside JavaScript using a generic, off-the-shelf sentence embedding model~\cite{SBERT}, demonstrating the ease of implementation of this interaction technique for arbitrary IEN games.

\section{Implementation}
The core of the PWIM interaction loop is as follows:

\begin{enumerate}
\item The player types a natural language phrase or sentence describing what they want to do.
\item Currently available game actions are sorted according to semantic similarity to this sentence.
\item Sorted actions are displayed in a way that visually prioritizes closer matches, but allows the player to choose what action is ultimately performed.
\end{enumerate}

In our demonstration, possible game actions are provided by a Praxish domain that mirrors the functionality of Versu. Praxish actions are defined as small blobs of structured JSON, making them relatively straightforward to author; see Fig. 3 in the original Praxish paper~\cite{Praxish} for an example.

Semantic similarity between available game actions and a player-entered intent phrase is determined via the \texttt{all-mpnet-base-v2} model provided with the open-source SentenceTransformers library~\cite{SBERT}. Every available action is summarized as a short imperative phrase, and embeddings of these action summary phrases are compared to an embedding of the player intent phrase via cosine similarity. Closer matches are displayed with a proportionally darker background color in the PWIM UI~(Fig. \ref{fig:PWIM}), and the top $K$ matches (in our demo, $K=3$) are displayed at an enlarged size to further increase their visual prevalence.

One way to explore the potential of this approach is to travel to the bar in our example game, then enter a wide variety of intent phrases. In our testing, we found that phrases as diverse as ``get hammered'', ``gimme something autumnal'', ``play music'', ``gotta stay hydrated'', and ``say hisaac'' all resulted in the most closely aligned possible action being surfaced as the top recommended action in the PWIM UI. Meanwhile, only a few testing phrases (e.g., ``sober up'') did not result in appropriate top recommendations. Occasional misclassification of intent phrases is why we choose to display all possible actions in the UI (giving players a chance to manually correct misclassifications) instead of automatically performing the action that represents the closest match; misclassifications can also likely be mitigated by finetuning the sentence embedding model against a game-specific set of player intent and action summary phrases, although this would increase deployment complexity.

\section{Related Work}
\citeauthor{TextToDialog}~\cite{TextToDialog} have previously employed a similar approach to ours, using sentence embeddings to map player utterances to pre-recorded lines of dialogue. However, this earlier work does not tie dialogue selection into a deeper emergent narrative simulation with complicated state tracking and many conditionally available actions. Additionally, player intent phrases are used in this work to automatically play the most closely matching line of dialogue; this simpler approach is more susceptible to derailment by misclassification of intent phrases, and may limit the discoverability of game actions in comparison to PWIM.

Some co-creative narrative play experiences have also made use of purely language modeling approaches---including data-driven case-based reasoning systems~\cite{SayAnything,CreativeHelp}, text-generating recurrent neural networks~\cite{CreativeHelpRNN}, and large language models~\cite{AIDungeon,1001Nights}---to respond to arbitrary text input by players. However, these systems generally do not attempt to ground out player input in a hand-crafted ontology of game actions, often limiting the coherence of the stories that these systems can produce---or, alternatively, requiring the intervention of an external game master to weave the results of command invocations into a coherent story~\cite{Avrae}.

\begin{acks}
Thanks to Barrett R. Anderson for initially exploring the possibility of moving the semantic similarity component of the PWIM demo into clientside JavaScript.
\end{acks}

\bibliographystyle{ACM-Reference-Format}
\bibliography{bibliography}


\end{document}